\documentclass[prb,twocolumn]{revtex4-1}
\usepackage{graphicx}
\usepackage{mathtools}
\usepackage{caption}
\usepackage{amsmath}
\usepackage{wasysym}
\usepackage{amsfonts}

\def\la{\left\langle}
\def\ra{\right\rangle}

\def\bfn{\hat{\bf n}}
\def\bfu{\hat{\bf u}}
\def\bfv{\hat{\bf v}}
\def\bfc{{\bf c}}
\def\bfj{{\bf j}}
\def\bfe{{\bf e}}
\def\bfr{{\bf r}}
\def\bfp{{\bf p}}
\def\bfz{{\bf z}}
\def\p{\partial }
\def\cross{\mbox{\boldmath $\times$}}
\def\amoon{a_{\mbox{\footnotesize\!\!\rightmoon}}}
\def\bfnmoon{\hat{\bf n}_{\mbox{\footnotesize\!\!\rightmoon}}}
\def\Mmoon{M_{\mbox{\footnotesize\!\!\rightmoon}}}
\def\Phimoon{\Phi_{\mbox{\footnotesize\!\!\rightmoon}}}
\def\betamoon{\beta_{\mbox{\footnotesize\!\!\rightmoon}}}

\def\rmoon{r_{\mbox{\footnotesize\!\!\rightmoon}}}
\def\bfrmoon{\bfr_{\mbox{\footnotesize\!\!\rightmoon}}}
\def\bnabla{\mbox{\boldmath $\nabla$}}
\def\llangle{\langle\langle}
\def\rrangle{\rangle\rangle}
\def\half{{\textstyle{1\over2}}}

\def\comment#1{}

\begin{document}

\title{Why do Earth satellites stay up?}

\author{ Scott Tremaine}

\email{tremaine@ias.edu}

\affiliation{Institute for Advanced Study,    Princeton, NJ 08540, USA}

\author{Tomer Yavetz}

\email{tyavetz@princeton.edu}

\affiliation{Department of Astrophysical Sciences, Peyton Hall, Princeton
University, Princeton, NJ 08540, USA}

\date{\today}

\begin{abstract}

\noindent
Satellites in low Earth orbits must accurately conserve their orbital
eccentricity, since a decrease in perigee of only 5--10\% would cause
them to crash. However, these satellites are subject to
gravitational perturbations from the Earth's multipole moments, the
Moon, and the Sun that are not spherically symmetric and hence do not
conserve angular momentum, especially over the tens of thousands of
orbits made by a typical satellite. Why then do satellites not crash?
We describe a vector-based analysis of the long-term behavior of
satellite orbits and apply this to several toy systems containing a
single non-Keplerian perturbing potential. If only the quadrupole or
$J_2$ potential from the Earth's equatorial bulge is present, all
near-circular orbits are stable. If only the octupole or $J_3$ potential is
present, all such orbits are unstable. If only the lunar or solar potential
is present, all near-circular orbits with inclinations to the ecliptic
exceeding $39^\circ$ are unstable. We describe the behavior of
satellites in the simultaneous presence of all of these perturbations
and show that almost all low Earth orbits are stable because of an
accidental property of the dominant quadrupole potential. We also relate
these results to the phenomenon of Lidov--Kozai oscillations.
\end{abstract}

\maketitle

\section{Introduction}
\label{sec:intro}

\noindent
The study of the motion of a satellite around its host planet is one
of the oldest problems in dynamics. This subject was re-invigorated at
the dawn of the Space Age in the late 1950s with a new focus on
artificial satellites in low Earth orbits.\citep{king-hele1958,el'yasberg1965}

Satellites orbiting a spherically symmetric Earth would have Keplerian
orbits, described by their semi-major axis $a$, eccentricity $e$,
inclination $\beta$, and other orbital elements. These orbits conserve the
energy $E$ and angular momentum ${\bf L}$ per unit mass, which for a
test particle are given by
\begin{equation}
E=-\frac{GM_\oplus}{2a}, \quad |{\bf L}|=[GM_\oplus a(1-e^2)]^{1/2}
\end{equation}
where $G$ is the gravitational constant and $M_\oplus$ is the mass of
the Earth. 
The perigee of the satellite orbit is $a(1-e)$, so if the energy and
angular momentum are conserved the perigee is conserved as well. This
is fortunate, since the typical perigee of satellites in low Earth
orbits is only $\sim 800\mbox{\;km}$ above the Earth. Orbits with
perigees less than 200--300 km are short-lived because of atmospheric
drag, so a decrease in perigee of less than 10\% would cause the satellite
orbit to decay quickly (we call this a ``crash'').

However, the Earth is not spherically symmetric. A better
approximation is an oblate spheroid, and in this potential the energy is
conserved but not the total angular momentum: if the polar axis of the
Earth is the $z$-axis, then only the $z$-component of the angular
momentum is conserved,
\begin{equation}
L_z=[GM_\oplus a(1-e^2)]^{1/2}\cos \beta.
\end{equation}
In this case the perigee is not necessarily conserved unless the orbit
lies in the Earth's equator. Why then do satellites in low Earth orbit
not crash?

A similar question arises if we treat the Earth as spherical and
consider the gravitational tides from the Moon and Sun, which are not
spherically symmetric and also do not conserve angular momentum.

Of course, the answer to why satellites do not crash is known, in the
sense that aerospace engineers design orbits that are stable over the
expected lifetime of the satellite. Nevertheless the calculations in
textbooks on astrodynamics and celestial mechanics do not provide a
physical explanation for this stability. In asking our colleagues we
have received a variety of answers, such as ``there are only periodic
oscillations in the perigee and these are too small to be important''
or ``the orbits are unstable but on timescales much longer than the
satellite lifetime''. We shall show that neither of these answers
captures the most relevant physics.

The primary goal of this paper is to understand physically the nature
of the strongest long-term perturbations to satellite orbits caused by
each non-spherical component that contributes to the gravitational
field around a planet. We will focus on understanding how these
components affect the stability of an initially circular orbit. We
have two secondary goals: first, we derive our results using a
geometric formalism that is simpler and more powerful for this purpose
than the usual algebraic methods used in celestial mechanics; second,
we show how our results are related to Lidov--Kozai oscillations, a
remarkable phenomenon in orbital mechanics that is relevant to a wide
variety of astrophysical systems.

\section{Secular dynamics of Earth satellites}

\subsection{The geometry of Keplerian orbits}

\noindent
We use a vectorial approach to calculate the behavior of the satellite
orbit. First we define a coordinate system whose equatorial plane
coincides with the satellite's orbital plane. The orthogonal unit
vectors of this system are $\bfn$ in the direction of the satellite's
angular-momentum vector, $\bfu$ pointing towards the perigee, and
$\bfv=\bfn\cross\bfu$. We introduce polar coordinates $(r,\phi)$ in
the orbital or $\bfu$--$\bfv$ plane, with $\phi=0$ coinciding with the $\bfu$ axis; then
the orbit of a test particle with semi-major axis $a$ and eccentricity
$e$ is given by
\begin{equation}
r=\frac{a(1-e^2)}{1+e\cos\phi},\qquad 
\bfr=r(\cos\phi\,\bfu+\sin\phi\,\bfv).
\label{eq:uv}
\end{equation}

The orbital period $P=2\pi a^{3/2}/(GM_\oplus)^{1/2}$ and the angular
momentum per unit mass $L=r^2 d\phi/dt=[(GM_\oplus
a)(1-e^2)]^{1/2}$. Using these results the average of some function
$\Phi(\bfr)$ over one orbit is
\begin{align}
\la\Phi(\bfr)\ra\equiv &\,\frac{1}{P}\int_0^P\!
dt\,\Phi[\bfr(t)] = \frac{(GM_\oplus)^{1/2}}{2\pi a^{3/2}}\int_0^{2\pi}\!\!
d\phi\,\frac{dt}{d\phi}\Phi(r,\phi) \nonumber \\
= &\,\frac{(GM_\oplus)^{1/2}}{2\pi
  a^{3/2}L}\int_0^{2\pi}\!\!
d\phi\,r^2\Phi(r,\phi)\nonumber \\
=&\,\frac{(1-e^2)^{3/2}}{2\pi} \int_0^{2\pi}\!\!
\frac{d\phi}{(1+e\cos\phi)^2}\,\Phi(r,\phi).
\label{eq:orbav}
\end{align}

The unit vectors $\bfn$ and $\bfu$ are
undefined for radial orbits ($e=1$) and circular orbits ($e=0$)
respectively. Therefore it is useful to introduce 
\begin{equation}
\bfj\equiv (1-e^2)^{1/2}\bfn, \quad \bfe=e\bfu
\label{eq:je}
\end{equation}
which are well-defined for all bound orbits. 
Apart from a factor $(GM_\oplus a)^{1/2}$, the first of these is the
angular-momentum vector per unit mass, and the second is the eccentricity or
Runge-Lenz vector. In terms of the position and momentum vectors,
\begin{equation}
\bfj=\frac{1}{(GM_\oplus a)^{1/2}}\,\bfr\cross\bfp, \quad
\bfe=\frac{1}{GM_\oplus}\bfp\cross(\bfr\cross\bfp)-\frac{\bfr}{r}.
\label{eq:jerv}
\end{equation}
We assume the satellite has unit mass so there is no distinction
between momentum and velocity.

\subsection{The multipole expansion of Earth's gravitational potential}
\label{sec:multipole}

\begin{table}[!h]
\centering
\caption{Constants for the Earth's gravitational
  potential and the solar and lunar tides\citep{wgs2000,luzum2011}}
\begin{ruledtabular}
\begin{tabular}{l l l}

constant & & value \\
\hline
Earth mass &$GM_\oplus$ & $3.9860\times10^{14}\ \text{m}^3\,\text{s}^{-2}$ \\
Earth radius &$R_\oplus$ & $6.3781\times10^6\ \text{m}$\\
Multipole moments& $J_2$ & $+1.0826\times10^{-3}$\\
 &$J_3$ & $-2.5327\times10^{-6}$\\
 &$J_4$ & $-1.6196\times10^{-6}$\\
 &$J_5$ & $-2.2730\times10^{-7}$\\
 &$J_6$ & $+5.4068\times10^{-7}$\\
 &$J_7$ & $-3.5236\times10^{-7}$\\
 &$J_8$ & $-2.0480\times10^{-7}$\\
solar mass &$GM_\odot$ & $1.327\times10^{20}\ \text{m}^3\,\text{s}^{-2}$ \\
Earth-Sun semi-major axis &$a_\odot$ & $1.496\times10^{11}\ \text{m}$\\
lunar mass &$G\Mmoon$ & $4.903\times10^{12}\ \text{m}^3\,\text{s}^{-2}$ \\
Earth-Moon semi-major axis &$\amoon$ & $3.844\times10^8\ \text{m}$

\end{tabular}
\end{ruledtabular}

\label{tab:multipole constants}

\end{table}

\noindent
The Earth's gravitational field can be represented as a multipole
expansion, of which the first several thousand terms have been
measured.\citep{wgs2000} Since our
goal is physical understanding, for simplicity we shall restrict
ourselves to the axisymmetric components of this field (the
  largest non-axisymmetric components are comparable to the $J_3$ and
  $J_4$ axisymmetric components in Eq.\  \ref{eq:multipole}). In this case
the potential can be written 
\begin{equation}
\label{eq:multipole} 
\Phi_\oplus(r,\theta) = -\frac{GM_\oplus}{r}\bigg[1-\sum_{n=2}^\infty J_n\left(\frac{R_\oplus}{r}\right)^nP_n(\cos\theta)\bigg] .
\end{equation}
Here $r$ and $\theta$ are spherical polar coordinates
relative to the Earth's spin axis $\bfn_\oplus$, $R_\oplus$ is the Earth radius, and
$P_n$ is a Legendre polynomial. The term with $n=1$ is zero if the
origin of the coordinates coincides with the Earth's center of
mass, as we shall assume. The values of these parameters, and of the first few multipole
moments $J_n$, are given in Table \ref{tab:multipole
  constants}. The first non-zero moment, $J_2$, is several
orders of magnitude larger than the others because of the equatorial bulge of
the Earth caused by its spin. 

We shall restrict our attention to the first three non-zero moments,
$J_2$, $J_3$, and $J_4$, since these are sufficient to illustrate how
a non-spherical potential affects satellite orbits. The methods we
describe are straightforward to extend to higher multipoles.

Since we are interested in small effects that accumulate over many
orbital times, we can average the gravitational potential over the
Keplerian orbit of a satellite with semi-major axis $a$ and
eccentricity $e$. We write the potential associated with $J_n$ in Eq.\ 
(\ref{eq:multipole}) as $\Phi_n$. Then, for example,
\begin{align}
\label{eq:J2} \Phi_2(r,\theta) =& \frac{GM_\oplus J_2
  R_\oplus^2}{r^3}P_2(\cos\theta) \nonumber \\
 =& \frac{GM_\oplus J_2 R_\oplus^2}{2}\left[3\frac{(\bfr\cdot\bfn_\oplus)^2}{r^5}-\frac{1}{r^3}\right]~.
\end{align}
Using Eq.\  (\ref{eq:uv}), the orbit average of $\Phi_2$ is
\begin{align}
\la\Phi_2\ra=&\; \frac{GM_\oplus J_2
  R_\oplus^2}{2}\bigg[3(\bfu\cdot\bfn_\oplus)^2\la\frac{\cos^2\phi}{r^3}\ra
+ 6(\bfu\cdot\bfn_\oplus)\nonumber \\ \times (\bfv\cdot\bfn_\oplus)\!&\la\frac{\cos\phi\sin\phi}{r^3}\ra\! +\!3(\bfv\cdot\bfn_\oplus)^2\la\frac{\sin^2\phi}{r^3}\ra\!-\!\la\frac{1}{r^3}\ra\!\bigg].
\end{align}
Using Eq.\  (\ref{eq:orbav}) we have
\begin{align}
\la\frac{\cos^2\phi}{r^3}\ra &= \la\frac{\sin^2\phi}{r^3}\ra
= \frac{1}{2a^3(1-e^2)^{3/2}}, \nonumber \\
\la\frac{\cos\phi\sin\phi}{r^3}\ra &= 0, \quad \la \frac{1}{r^3}\ra=\frac{1}{a^3(1-e^2)^{3/2}}.
\end{align}
Finally, since $(\bfu,\bfv,\bfn)$ is an orthonormal triad, 
$(\bfu\cdot\bfn_\oplus)^2+(\bfv\cdot\bfn_\oplus)^2+(\bfn\cdot\bfn_\oplus)^2=1$,
and this can be used to eliminate $(\bfu\cdot\bfn_\oplus)^2$ and
$(\bfv\cdot\bfn_\oplus)^2$. Thus\comment{I have checked this
  against Kozai 1959}
\begin{equation}
\label{} \la\Phi_2\ra =  \frac{GM_\oplus J_2 R_\oplus^2}{4a^3(1-e^2)^{3/2}}\left[1-3(\bfn_\oplus\cdot\bfn)^2\right] .
\end{equation}
In terms of the vectors $\bfj$ and $\bfe$ (Eq.\ \ref{eq:je}) 
\begin{equation}
\label{eq:quadpot} \la\Phi_2\ra =  \frac{GM_\oplus J_2 R_\oplus^2}{4a^3(1-e^2)^{5/2}}\left[1-e^2-3(\bfj\cdot\bfn_\oplus)^2\right]\, .
\end{equation}

Similarly, \comment{I have checked both of these against Kozai 1959}
\begin{align}
\la\Phi_3\ra =&\;\frac{3GM_\oplus J_3
  R_\oplus^3}{8a^4(1-e^2)^{7/2}}(\bfe\cdot\bfn_\oplus)
\left[1-e^2-5(\bfj\cdot\bfn_\oplus)^2\right] ; \nonumber \\
\la\Phi_4\ra = &\;\frac{3GM_\oplus J_4
  R_\oplus^4}{128a^5(1-e^2)^{11/2}}\Big\{(6-e^2)(1-e^2)^2 \nonumber \\
-10&(6+e^2)(1-e^2)(\bfj\cdot\bfn_\oplus)^2+35(2+e^2)(\bfj\cdot\bfn_\oplus)^4\nonumber
\\
+20&(\bfe\cdot\bfn_\oplus)^2
(1-e^2)\left[1-e^2-7(\bfj\cdot\bfn_\oplus)^2\right]\Big\} ~.
\label{eq:phithreefour}
\end{align}
Since $j^2+e^2=1$ any terms involving $e^2$ can be replaced by
$1-j^2$. 

\subsection{The gravitational potential from the Moon}

\noindent
The effects on satellite orbits of the tides from the Sun and the Moon
are qualitatively similar, and since the lunar tide is stronger by a
factor of about 2.2 we consider only the Moon. 

The gravitational potential from the Moon is
$\Phimoon(\bfr,\bfrmoon)=-G\Mmoon/|\bfr-\bfrmoon|$. Since
$|\bfr|\ll|\bfrmoon|$ (by a factor of about 60), we expand this
  potential as a Taylor series,
\begin{equation}
\Phimoon(\bfr,\bfrmoon)=-\frac{G\Mmoon}{\rmoon}\left[1+\frac{\bfr\cdot\bfrmoon}{\rmoon^2}-\frac{r^2}{2\rmoon^2}
+\frac{3(\bfr\cdot\bfrmoon)^2}{2\rmoon^4}\right]
\end{equation}
plus terms of order $(r/\rmoon)^3$ and higher, which we drop. The
first term is independent of the dynamical variable $\bfr$ and so can
also be dropped; the second term is canceled by the fictitious
potential due to the acceleration of the Earth's center of mass by the
Moon. Then averaging over the satellite orbit using equation
(\ref{eq:uv}) we have
\begin{align}
&\la\Phimoon\ra =\frac{G\Mmoon}{2\rmoon^3}\bigg[\la
  r^2\ra -\frac{3(\bfu\cdot\bfrmoon)^2}{\rmoon^2}\la
  r^2\cos^2\!\phi\ra \nonumber \\
&-\frac{6(\bfu\cdot\bfrmoon)(\bfv\cdot\bfrmoon)}{\rmoon^2}\!\la
  r^2\!\cos\phi\sin\phi\ra -\frac{3(\bfv\cdot\bfrmoon)^2}{\rmoon^2}\la
  r^2\!\sin^2\!\phi\ra\!\bigg].
\end{align}
Using Eq.\  (\ref{eq:orbav}) we have
\begin{align}
\la r^2\cos^2\phi\ra &= \frac{a^2}{2}(1+4e^2), \quad 
\la r^2\cos\phi\sin\phi\ra = 0, \nonumber \\ \la r^2\sin^2\phi\ra &=\frac{a^2}{2}(1-e^2),
\quad \la r^2\ra=\frac{a^2}{2}(2+3e^2).
\end{align}
Finally, we eliminate $(\bfv\cdot\bfrmoon)^2$ using the relation
$\rmoon^2=(\bfu\cdot\bfrmoon)^2+(\bfv\cdot\bfrmoon)^2+(\bfn\cdot\bfrmoon)^2$
and replace $\bfu$ and $\bfn$ with $\bfj$ and $\bfe$ using equation
(\ref{eq:je}):
\begin{equation}
\la\Phimoon\ra=\frac{G\Mmoon a^2}{4\rmoon^3}\Big[-1
  +6e^2 +\frac{3(\bfj\cdot\bfrmoon)^2}{\rmoon^2}
  -\frac{15(\bfe\cdot\bfrmoon)^2}{\rmoon^2}\Big]. 
\end{equation}

For simplicity we assume that the Moon's orbit is circular, with
semi-major axis $\amoon$ and a fixed normal $\bfnmoon$. Averaging any
fixed vector $\bfc$ over the Moon's orbit, we have
$\la(\bfc\cdot\bfrmoon)^2\ra=\frac{1}{2}\amoon^2[c^2-(\bfc\cdot\bfnmoon)^2]$. Thus
\begin{equation}
\label{eq:ejdef} \llangle\Phimoon\rrangle =
\frac{G\Mmoon a^2}{8\amoon^3}\Big[15(\bfe\cdot\bfnmoon)^2
  -3(\bfj\cdot\bfnmoon)^2+1-6e^2\Big]~.
\end{equation}

\subsection{Rotating reference frame}

\noindent
In some cases it will be useful to work in a reference frame that
rotates about the Earth's spin axis with angular speed $\omega$. In
this frame the Lagrangian for a particle of unit mass in a potential $\Phi(\bfr)$ is
$L=\frac{1}{2}(\dot\bfr+\omega\,
\bfn_\oplus\cross\bfr)^2-\Phi(\bfr)$. The canonical momentum is
$\bfp=\p L/\p \dot\bfr=\dot\bfr+\omega\,
\bfn_\oplus\cross\bfr$. The Hamiltonian is
$H=\bfp\cdot\dot\bfr-L=\frac{1}{2}\bfp^2+\Phi(\bfr)-\omega\,\bfp\cdot\bfn_\oplus\cross\bfr=\frac{1}{2}\bfp^2+\Phi(\bfr)-\omega\,\bfn_\oplus\cdot\bfr\cross\bfp
=\frac{1}{2}\bfp^2+\Phi(\bfr)-\omega(GM_\oplus
a)^{1/2}\bfj\cdot\bfn_\oplus$, where the last equality follows from
(\ref{eq:jerv}). Thus transforming to a rotating frame simply adds a
term 
\begin{equation}
\Phi_{\rm rot}=-\omega(GM_\oplus a)^{1/2}\,\bfj\cdot\bfn_\oplus
\label{eq:rot}
\end{equation}
to the Hamiltonian. 

\subsection{Equations of motion}
\label{sec:secular multipole}

\noindent
Having derived the orbit-averaged perturbing potentials from the
Earth's multipole moments and the lunar tidal field, we must now
determine how the satellite orbit responds to these
perturbations. Once again we use a vectorial method. 

The evolution of the orbit is determined by a Hamiltonian $H=H_{\rm
  Kep}+\Phi(\bfr)$, where $H_{\rm Kep}=\frac{1}{2}v^2-GM_\oplus/r$ is
the Kepler Hamiltonian. If $\Phi(\bfr)=0$ the energy $E_{\rm
  Kep}=-\frac{1}{2}GM_\oplus/a$ is conserved, where as usual $a$ is
the semi-major axis. If the perturbing potential is non-zero, $dE_{\rm
  Kep}/dt=\bfp\cdot\bnabla\Phi$. If we orbit-average, as we did for
the potentials, $\la\bfp\cdot\bnabla\Phi\ra=0$ so $E_{\rm Kep}$ and
the semi-major axis are conserved. Then the orbit's shape and
orientation are determined entirely by the vectors $\bfj$ and $\bfe$
(Eq.\ \ref{eq:je}).

From Eq.\  (\ref{eq:jerv}) the Poisson brackets of $\bfj$ and
$\bfe$ are
\begin{align}
\{j_i,j_j\}=&\frac{1}{\sqrt{GM_\oplus a}}\epsilon_{ijk}j_k, \quad 
\{e_i,e_j\}=\frac{1}{\sqrt{GM_\oplus a}}\epsilon_{ijk}j_k, \nonumber \\
\{j_i,e_j\}=&\frac{1}{\sqrt{GM_\oplus a}}\epsilon_{ijk}e_k.
\label{eq:pois}
\end{align}
The time evolution under the Hamiltonian $H$ of any function $f$ of
the phase-space coordinates is
\begin{equation}
\frac{df}{dt}=\{f,H\}.
\end{equation}
Then by the chain rule
\begin{equation}
\frac{df}{dt}=\{f,\bfj\}\bnabla_\bfj H + \{f,\bfe\}\bnabla_\bfe H,
\end{equation}
where $\bnabla_\bfj$ is the vector $(\p/\p j_1,\p/\p j_2,\p/\p j_3)$,
and similarly for $\bnabla_\bfe$. Replacing $f$ by $j_i$ and $e_i$ and
using relations (\ref{eq:pois}) we have \citep{mil39}
\begin{align}
\frac{d\bfj}{dt}=-\,\frac{1}{\sqrt{GM_\oplus
    a}}\left(\bfj\cross\bnabla_\bfj
  H + \bfe\cross\bnabla_\bfe H\right), \nonumber \\
\frac{d\bfe}{dt}=-\,\frac{1}{\sqrt{GM_\oplus
    a}}\left(\bfj\cross\bnabla_\bfe H + \bfe\cross\bnabla_\bfj
  H\right).
\label{eq:mot}
\end{align}
Since $\bfe$ and $\bfj$ are constants of motion for the Kepler
Hamiltonian $H_{\rm Kep}$, we can replace $H=H_{\rm Kep} +\la\Phi\ra$
by $\la\Phi\ra$. 

By definition, $\bfj$ and $\bfe$ must satisfy:
\begin{equation}
\label{ej relations} \bfj\cdot\bfe = 0~,~~~~~ \bfj^2+\bfe^2 = 1 ~.
\end{equation}
It is straightforward to show that these constraints are conserved by
the equations of motion (\ref{eq:mot}). There is a gauge freedom in
the Hamiltonian $H$ since it can be replaced by $H+F$ where
$F(\bfj,\bfe)$ is any function that is constant on the manifold
(\ref{ej relations}); however, adding this function has no effect on
the equations of motion.\citep{ttn09}

\section{Stability of satellites on circular orbits}

\noindent
We now investigate the effect of each of the perturbations we have
discussed---the multipole moments $J_2$, $J_3$, and $J_4$ and the
lunar tide---on the stability of a satellite on a circular low Earth
orbit. We have verified most of these conclusions by numerical
integrations of test-particle orbits.

\subsection{Quadrupole ($\mathbf{J_2}$) potential} 

\noindent
If the only perturbation is the quadrupole term (\ref{eq:quadpot}) we
have
\begin{align}
\bnabla_\bfj \la\Phi_2\ra & =
-\frac{3GM_\oplus J_2R_\oplus^2}{2a^3(1-e^2)^{5/2}}(\bfj\cdot\bfn_\oplus)\bfn_\oplus
~, \nonumber \\
\label{eq:grade2} \bnabla_\bfe \la\Phi_2\ra & =  \frac{3GM_\oplus J_2R_\oplus^2}{4a^3(1-e^2)^{7/2}}\left[1-e^2-5(\bfj\cdot\bfn_\oplus)^2\right]\bfe~.
\end{align}
Substituting these results into the equations of motion (\ref{eq:mot})
and dropping all terms higher than linear order in eccentricity,
\begin{align}
\frac{d\bfj}{dt} &= \frac{3(GM_\oplus)^{1/2}
  J_2R_\oplus^2}{2a^{7/2}}(\bfj\cdot\bfn_\oplus)\,\bfj\cross\bfn_\oplus,
\nonumber \\
\frac{d\bfe}{dt} &= \frac{3(GM_\oplus)^{1/2}
  J_2R_\oplus^2}{4a^{7/2}}\Big\{
  \left[1-5(\bfj\cdot\bfn_\oplus)^2\right]\,\bfe\cross\bfj \nonumber
  \\
&\qquad + 2(\bfj\cdot\bfn_\oplus)\,\bfe\cross\bfn_\oplus\Big\}.
\end{align}
The first of these equations describes precession of the satellite's
orbital angular momentum $\bfj$ around the symmetry axis of the Earth
$\bfn_\oplus$. The angular frequency of the precession is 
\begin{equation}
\omega=-\frac{3(GM_\oplus)^{1/2}
  J_2R_\oplus^2}{2a^{7/2}}(\bfj\cdot\bfn_\oplus)
\label{eq:omdef}
\end{equation}
where $\bfj\cdot\bfn_\oplus=\cos \beta$ and $\beta$ is the constant inclination of the
orbit to the Earth's equator. The minus sign indicates retrograde precession
of the angular momentum vector around the symmetry axis. 

The second equation is best analyzed by transforming to the frame
rotating with the precession of the orbit. In this frame $\bfj$ is a
constant; the Hamiltonian is modified according to equation
(\ref{eq:rot}); and the equation of motion for the eccentricity
is modified to
\begin{equation}
\frac{d\bfe}{dt} = \frac{\omega}{2\cos \beta}\left(5\cos^2\beta-1\right)\,\bfe\cross\bfj.
\end{equation}
We now switch to Cartesian coordinates with the positive $z$-axis
along $\bfn_\oplus$ and with $\bfj$ in the $x$-$z$ plane, so
$\bfj=(\sin \beta,0,\cos \beta)$. We look for a solution of the form
$\bfe=\bfe_0\exp(\lambda t)$; solving this simple eigenvalue problem
yields either $\lambda=0$ or 
\begin{align}
\lambda=&\;\pm i
\frac{\omega}{2\cos\beta}(1-5\cos^2\beta) \nonumber \\ =&\; \mp i \frac{3(GM_\oplus)^{1/2}
  J_2R_\oplus^2}{4a^{7/2}}(1-5\cos^2\beta).
\label{eq:lambda}
\end{align}

Since the eigenvalues are zero or purely imaginary, we conclude that nearly
circular low Earth orbits of all inclinations are stable under the influence of the quadrupole
moment of the Earth. Note that if the inclination $\beta$ satisfies
$1-5\cos^2{\beta} = 0$, or $\beta = \beta_{\rm crit}=\cos^{-1}\sqrt{1/5}=63.43^\circ$, all three
eigenvalues $\lambda$ are zero in Eq.~(\ref{eq:lambda}). This is 
the so-called \emph{critical inclination}, and the motion of satellites in the
vicinity of this inclination has been the subject of many
studies.\citep{crit}

In physical units, the timescale of these oscillations is: 
\begin{equation}
\label{}
|\lambda|^{-1}=\frac{11.5\mbox{\,d}}{|1-5\cos^2\beta|}\left(\frac{a}{R_\oplus}\right)^{7/2}\!.
\end{equation}

\subsection{Octupole ($\mathbf{J_3}$) potential}

\noindent
If $J_3$ is the only non-zero multipole moment then Eq.\ 
(\ref{eq:phithreefour}) gives
\begin{align}
\bnabla_\bfj \la\Phi_3\ra & =
-\frac{15GM_\oplus J_3R_\oplus^3}{4a^4(1-e^2)^{7/2}}(\bfj\cdot\bfn_\oplus)(\bfe\cdot\bfn_\oplus)\bfn_\oplus
~, \nonumber \\
\bnabla_\bfe \la\Phi_3\ra & =  \frac{3GM_\oplus
  J_3R_\oplus^3}{8a^4(1-e^2)^{7/2}}\Big\{\bfn_\oplus\left[1-e^2-5(\bfj\cdot\bfn_\oplus)^2\right]
\nonumber \\
  + \,5&\frac{\bfe}{1-e^2}(\bfe\cdot\bfn_\oplus)\left[1-e^2-7(\bfj\cdot\bfn_\oplus)^2\right]\Big\}.
\end{align}
We substitute these results into the equations of motion
(\ref{eq:mot}); in this case stability can be determined by dropping
all terms linear or higher in the eccentricity from the right-hand
side and we have 
\begin{align}
\frac{d\bfj}{dt} &= 0, 
\nonumber \\
\frac{d\bfe}{dt} &= -\frac{3(GM_\oplus)^{1/2}
  J_3R_\oplus^3}{8a^{9/2}}\left[1-5(\bfj\cdot\bfn_\oplus)^2\right]\,\bfj\cross\bfn_\oplus.
\label{eq:linear}
\end{align}
Thus the eccentricity vector grows linearly with time, $\bfe(t)=\bfe_0
+ \bfc t$ where
\begin{equation}
|\bfc|=\frac{3(GM_\oplus)^{1/2}
  |J_3|R_\oplus^3}{8a^{9/2}}\sin\beta\left(1-5\cos^2\beta\right).
\end{equation}
We conclude that the $J_3$ term causes instability for all
inclinations except $\beta=0$ (equatorial orbit) and $\beta=\beta_{\rm
  crit}$. Hence if only the $J_3$ multipole were non-zero and there
were no other external forces, Earth satellites on all orbits except
these two would crash. The survival timescale would be much smaller
than
\begin{equation}
|\bfc|^{-1}=\frac{26.9\mbox{\,yr}}{\sin\beta|1-5\cos^2\beta|}\left(\frac{a}{R_\oplus}\right)^{9/2}.
\end{equation}

\subsection{$\mathbf{J_4}$ potential} 

\label{sec:jfour}

\noindent
The analysis of orbital stability for higher
multipoles proceeds in the same way as for $J_2$ and $J_3$, although
the algebra rapidly becomes more complicated. For $J_4$, we substitute
the averaged potential $\la\Phi_4\ra$ (Eq.\ \ref{eq:phithreefour})
into the equations of motion (\ref{eq:mot}). Keeping terms up to first
order in eccentricity on the right-hand side we find 
\begin{align}
\frac{d\bfj}{dt} = &
\frac{15(GM_\oplus)^{1/2}J_4R_\oplus^4}{16a^{11/2}}\left[3(\bfj\cdot\bfn_\oplus)-
7(\bfj\cdot\bfn_\oplus)^3\right]\,\bfj\cross\bfn_\oplus ,
\nonumber \\
\frac{d\bfe}{dt} = &
\frac{15(GM_\oplus)^{1/2}J_4R_\oplus^4}{16a^{11/2}}\Big\{\!\left[3(\bfj\cdot\bfn_\oplus)-
7(\bfj\cdot\bfn_\oplus)^3\right]\bfe\cross\bfn_\oplus\nonumber \\
& \quad -\left[1-7(\bfj\cdot\bfn_\oplus)^2\right](\bfe\cdot\bfn_\oplus)\,\bfj\cross\bfn_\oplus 
\nonumber\\
&\quad -\left[1-14(\bfj\cdot\bfn_\oplus)^2+ 21(\bfj\cdot\bfn_\oplus)^4\right]\,\bfj\cross\bfe\Big\} ~.
\end{align}
The first of these equations describes precession of the satellite's
orbital angular momentum, with angular frequency
\begin{align}
\omega=&\;-\frac{15(GM_\oplus)^{1/2}J_4R_\oplus^4}{16a^{11/2}}\left[3(\bfj\cdot\bfn_\oplus)-
    7(\bfj\cdot\bfn_\oplus)^3\right] \nonumber \\=&\;
-\frac{15(GM_\oplus)^{1/2}J_4R_\oplus^4}{16a^{11/2}}\left(3\cos\beta-7\cos^3\beta\right),
\end{align}
where the inclination $\beta$ is a constant.  The second equation is
best analyzed by transforming to the frame rotating with the
precession of the orbit. In this frame $\bfj$ is a constant; the
Hamiltonian is modified according to Eq.\  (\ref{eq:rot}); and the
equation of motion for the eccentricity (\ref{eq:mot}) is modified to
\begin{align}
\frac{d\bfe}{dt} =&\, -\frac{15(GM_\oplus)^{1/2}J_4R_\oplus^4}{16a^{11/2}}\big\{\!
\left(1-7\cos^2\beta\right)
(\bfe\cdot\bfn_\oplus)\,\bfj\cross\bfn_\oplus \nonumber \\
&\qquad+\left(1-14\cos^2\beta+ 21\cos^4\beta\right)\,\bfj\cross\bfe \big\}~.
\end{align}
We assume $\bfe=\bfe_0\exp(\lambda t)$ and solve the resulting
eigenvalue equation for $\lambda$. We find that either $\lambda=0$ or 
\begin{align}
\lambda=&\;\pm
i\frac{15(GM_\oplus)^{1/2}J_4R_\oplus^4}{64a^{11/2}}(3+6\cos2\beta+7\cos4\beta)^{1/2}
\nonumber \\ &\qquad \times(15+28\cos2\beta+21\cos4\beta)^{1/2} ~.
\end{align}

For nearly all choices of $\beta$, the value of $\lambda$ is purely
imaginary, indicating a stable orbit. However, for $34.40^\circ <
\beta < 40.09^\circ$, and for $71.10^\circ < \beta < 73.42^\circ$, the orbits are
unstable.  Taking the inclination 
$37.04^\circ$ that maximizes the growth rate, the characteristic timescale for
this growth is
\begin{equation}
\label{eq:predict}  |\lambda|^{-1}= 26.8\mbox{\;yr}\left(\frac{a}{R_\oplus}\right)^{11/2}.
\end{equation}

For the second unstable regime ($71.10^\circ < \beta < 73.40^\circ$),
the maximum growth rate occurs at $\beta=72.26^\circ$ and the
characteristic growth time is larger by a factor of 4.08. 

\subsection{The lunar potential} 

\label{sec:moon}

\noindent
We substitute the doubly averaged potential
$\llangle\Phimoon\rrangle$ from Eq.\  (\ref{eq:ejdef}) into the equations of motion (\ref{eq:mot}):
\begin{align}
\frac{d\bfj}{dt} = & \;\frac{3G^{1/2}\Mmoon a^{3/2}}{4M_\oplus^{1/2}\amoon^3}\big[(\bfj\cdot
  \bfnmoon)\,\bfj\cross\bfnmoon-5(\bfe\cdot\bfnmoon)\,\bfe\cross\bfnmoon\big], \nonumber \\
\label{eq:dedto} \frac{d\bfe}{dt} = &\;
\frac{3G^{1/2}\Mmoon a^{3/2}}{4M_\oplus^{1/2}\amoon^3}\big[(\bfj\cdot
  \bfnmoon)\,\bfe\cross\bfnmoon-5(\bfe\cdot\bfnmoon)\,\bfj\cross\bfnmoon
  \nonumber \\
&\qquad +2\,\bfj\cross\bfe\big]~.
\end{align}
To first order in the eccentricity, the first of these 
describes uniform precession of $\bfj$ around $\bfnmoon$ with
frequency
\begin{equation}
\omega=-\frac{3(GM_\oplus)^{1/2}a^{3/2}}{4\amoon^3}\frac{\Mmoon}{M_\oplus}(\bfj\cdot
  \bfnmoon),
\end{equation}
where $\bfj\cdot\bfnmoon=\cos\betamoon$ and $\betamoon$ is the
inclination of the satellite orbit to the lunar orbit, which is nearly
in the ecliptic plane. We then
transform to the frame rotating at this frequency and the equation of
motion for the eccentricity simplifies to
\begin{equation}
\frac{d\bfe}{dt} = 
\frac{3(GM_\oplus)^{1/2}a^{3/2}}{4\amoon^3}\frac{\Mmoon}{M_\oplus}\big[2\,\bfj\cross\bfe-5(\bfe\cdot\bfnmoon)\,\bfj\cross\bfnmoon\big]~.
\end{equation}
Assuming $\bfe=\bfe_0\exp(\lambda t)$ and solving the eigenvalue
equation, we find $\lambda=0$ or 
\begin{equation}
\lambda=\pm\frac{3(GM_\oplus)^{1/2}a^{3/2}}{2^{3/2}\amoon^3}\frac{\Mmoon}{M_\oplus}(3-5\cos^2\betamoon)^{1/2}.
\label{eq:betacrit}
\end{equation}
The solution is stable if and only if the inclination
$\betamoon<\beta_{\rm L}\equiv \cos^{-1}\sqrt{3/5}=39.23^\circ$ \citep{lid61} or
$\betamoon>180^\circ-\beta_{\rm L}$. In other words, all nearly
circular orbits with an inclination between $39.23^\circ$ and
$140.77^\circ$ with respect to the lunar orbit will have an exponentially increasing eccentricity.

The growth rate is 
\begin{equation}
|\lambda|^{-1}=\frac{248\mbox{\;yr}}{\sqrt{1-\frac{5}{3}\cos^2\betamoon}}\left(\frac{R_\oplus}{a}\right)^{3/2}.
\end{equation}

\subsection{Power-law quadrupole potential}

\noindent
It is striking that both the $J_2$ potential and the lunar potential
are quadrupoles [angular dependence $\propto P_2(\cos\theta)]$, yet
circular orbits are stable in one of these perturbing potentials at
all inclinations and unstable in the other for a wide range of
inclinations. Some insight into this behavior comes from examining an
artificial axisymmetric perturbing potential
$\Phi_c=cr^bP_2(\cos\theta)$ with symmetry axis $\bfn_c$; the $J_2$
potential corresponds to $b=-3$ and the lunar potential to 
$b=+2$. After orbit-averaging and dropping all terms of order higher
than $e^2$,
\begin{align}
&\la\Phi_c\ra=\frac{ca^b}{32}\big\{6(2+b)(3+b)(\bfe\cdot\bfn_c)^2+8
\nonumber \\
&\ -(18+13b+b^2)e^2
  - 3[8+(2-3b+b^2)e^2](\bfj\cdot\bfn_c)^2\big\}.
\end{align}
We substitute this potential into the equations of motion
(\ref{eq:mot}) and find $\bfe=\bfe_0\exp(\lambda t)$ with
\begin{align}
\lambda&=\pm
\frac{ca^b}{16\sqrt{GM_\oplus a}}\nonumber \\
&\times\bigg\{\big[18+13b+b^2+3(2-3b+b^2)\cos^2\beta\big]
  \nonumber \\
&\times \big[18+17b+5b^2-3(14+7b+3b^2)\cos^2\beta\big]\bigg\}^{1/2}
\end{align}
where $\cos\beta=\bfn\cdot\bfn_c$.  The satellite orbit is unstable if
the quantity in braces is positive.

\begin{figure}[h!]
\centering
\includegraphics[width=0.95\hsize]{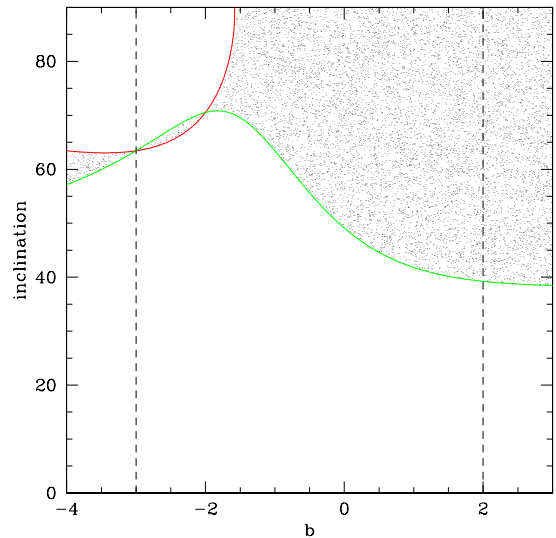}
\vspace{-2cm}
\caption{Stability of nearly circular orbits in a potential of the
  form $r^bP_2(\cos\theta)$. The horizontal axis is the exponent $b$
  and the vertical axis is the orbit inclination relative to the
  symmetry plane of the potential. Unstable regions of parameter space
  are stippled, and bounded by red and green lines. Vacuum solutions
  of Poisson's equations require $b=-3$ or $2$, and are marked by
  vertical dashed lines.}
\label{fig:one}
\end{figure}

Figure \ref{fig:one} shows the unstable inclinations (stippled) as a
function of the inclination $\beta$ and the exponent of the potential,
$b$. Remarkably, the only exponents for which all inclinations are
stable are $b=-3$, corresponding to the potential arising from the
planetary quadrupole $J_2$, and $b=-2$, which is unphysical. We
conclude that the stability of orbits of all inclinations circling an
oblate planet is an accidental consequence of the properties of the
quadrupole potential. 

\subsection{General axisymmetric potential} 

\noindent
Now consider the more
general case of an arbitrary axisymmetric perturbation $\Phi(\bfr)$ to
the Kepler potential. Let $\hat\bfz$ be the symmetry axis of the
potential. The orbit-averaged potential $\la\Phi\ra$ can only depend
on the semi-major axis $a$ and the dimensionless eccentricity and
angular-momentum vectors $\bfe$, $\bfj$. Since the semi-major axis is
fixed in the orbit-averaged equations of motion (\S\ref{sec:secular
  multipole}) we do not need to consider the dependence of
$\la\Phi\ra$ on it. We shall assume that the potential $\Phi(\bfr)$ is
a smooth function of position $\bfr$, so $\la\Phi\ra$ must be a smooth
function of $\bfe$ and $\bfj$. Moreover because of axisymmetry the
dependence on these vectors can only occur through the scalars $e^2$,
$j^2$, $\bfe\cdot\hat\bfz$ and $\bfj\cdot\hat\bfz$; and since
$j^2+e^2=1$ we need only one of the first two in this list. Finally, since
the shape of the orbit is unchanged if $\bfj\to-\bfj$ the potential
can only depend on $(\bfj\cdot\hat\bfz)^2$.  Thus the orbit-averaged
potential can be written
$\la\Phi\ra[e^2,\bfe\cdot\hat\bfz,(\bfj\cdot\hat\bfz)^2]$.  Denoting
the derivative with respect to argument $i$ by $\Phi_{,i}$ the equations
of motion (\ref{eq:mot}) become
\begin{align}
   \frac{d\bfj}{dt}&=-\frac{1}{\sqrt{GMa}}\left[\la\Phi\ra_{,2}\,\bfe\cross\hat\bfz
                   +2 \la\Phi\ra_{,3}(\bfj\cdot\hat\bfz)\, \bfj\cross\hat\bfz\right]  \nonumber \\
   \frac{d\bfe}{
     dt}&=-\frac{1}{\sqrt{GMa}}\Big[2\la\Phi\ra_{,1}\,\bfj\cross\bfe+\la\Phi\ra_{,2}\,\bfj\cross\hat\bfz
     \nonumber \\ & \qquad\quad + 2 \la\Phi\ra_{,3}  (\bfj\cdot\hat\bfz)\,\bfe\cross\hat\bfz)\Big]. 
\label{eq:eqmota}
\end{align}
The first of these shows that $\bfj\cdot\hat\bfz$ is constant, i.e., the
$z$-component of angular momentum is conserved. 

Now assume $|\bfe|\ll1$ and keep only the terms on the right side that are
independent of $\bfe$ in the first equation and up to linear in $\bfe$ in the second:
\begin{align}
   {d\bfj\over dt}&=-{2\over\sqrt{GMa}}\la\Phi\ra_{,3}  (\bfj\cdot\hat\bfz) \,\bfj\cross\hat\bfz \nonumber \\
   {d\bfe\over
     dt}&=-{1\over\sqrt{GMa}}\Big[2\la\Phi\ra_{,1}\,\bfj\cross\bfe+\la\Phi\ra_{,2}\,\bfj\cross\hat\bfz
\nonumber \\
     &\quad + \la\Phi\ra_{,22} (\bfe\cdot\hat\bfz)\,\bfj\cross\hat\bfz+ 2 \la\Phi\ra_{,3} (\bfj\cdot\hat\bfz) \,\bfe\cross\hat\bfz\Big]. 
\label{eq:eqmotb}
\end{align}
where the derivatives of $\la\Phi\ra$ are evaluated at $e=0$. Since $\bfj\cdot\hat\bfz$ is
constant, these derivatives can be taken to be constants. The first equation describes
uniform precession of $\bfj$ around $\hat\bfz$ at angular frequency 
\begin{equation}
\omega={2\la\Phi\ra_{,3}\over\sqrt{GMa}}(\bfj\cdot\hat\bfz).
\end{equation}
In the frame rotating at $\omega$, $d\bfj/dt=0$; the Hamiltonian is
modified according to Eq.\  (\ref{eq:rot}); and the equation of
motion for the eccentricity is modified to
\begin{align}
   {d\bfe\over
     dt}=&\;-{1\over\sqrt{GMa}}\Big[2\la\Phi\ra_{,1}\,\bfj\cross\bfe+\la\Phi\ra_{,2}\,\bfj\cross\hat\bfz
\nonumber \\ 
     &\qquad + \la\Phi\ra_{,22} (\bfe\cdot\hat\bfz)\,\bfj\cross\hat\bfz\Big]. 
\label{eq:hhh}
\end{align}
All quantities other than $\bfe$ on the right side are constants, so
the general solution is of the form $\bfe=\bfe_0+\sum_{i=1}^2\bfc_i\exp(\lambda_i t)$ where\cite{kd11}
\begin{equation}
\lambda=\pm
{2i\over\sqrt{GMa}}\left[\la\Phi\ra_{,1}\big(\la\Phi\ra_{,1}+\half\la\Phi\ra_{,22}\sin^2\beta\big)\right]^{1/2};
\label{eq:ggg}
\end{equation}
\begin{equation}
2\la\Phi\ra_{,1}\,\bfj\cross\bfe_0+ \la\Phi\ra_{,22}
(\bfe_0\cdot\hat\bfz)\,\bfj\cross\hat\bfz=-\la\Phi\ra_{,2}\,\bfj\cross\hat\bfz,
\label{eq:kkk}
\end{equation}
and $\cos \beta=\bfj\cdot\hat\bfz$ for circular orbits.

With this formula several of our earlier results become easy to
interpret:

\noindent (i) The potential
$\la\Phi_2\ra[e^2,\bfe\cdot\hat\bfz,(\bfj\cdot\hat\bfz)^2]$ associated
with the Earth's quadrupole moment (Eq.\ \ref{eq:quadpot}) has no
dependence on $\bfe\cdot\bfn_\oplus$ so $\la\Phi_2\ra_{,22}=0$ and
$\lambda$ is always pure imaginary, which explains why orbits in this
perturbing potential are always stable.

\noindent (ii) The potential
$\la\Phi_3\ra[e^2,\bfe\cdot\hat\bfz,(\bfj\cdot\hat\bfz)^2]$ associated
with the multipole moment $J_3$ is an odd function of
$\bfe\cdot\hat\bfz$ (Eq.\ \ref{eq:phithreefour}). Thus
$\la\Phi_3\ra_{,1}$ is also odd, so it is zero when evaluated at
$e=0$. Then Eq.\  (\ref{eq:ggg}) implies that both eigenvalues
$\lambda$ are zero. Two degenerate eigenvalues give rise to linear
growth in the solution to equations like (\ref{eq:hhh}), which is what
was seen in the solution to Eq.\  (\ref{eq:linear}). 

\noindent (iii) Note also that $\la\Phi_3\ra$ is linear in
$\bfe\cdot\bfz$ (Eq.\ \ref{eq:phithreefour}) so
$\la\Phi_3\ra_{,22}=0$. Thus for the combined potential
$\la\Phi_2+\Phi_3\ra$, Eq.\ (\ref{eq:ggg}) is simply $\lambda=\pm
2i\la\Phi_2\ra_{,1}/\sqrt{GMa}$, independent of $\Phi_3$. We conclude
that any quadrupole potential, no matter how small, will stabilize a
circular satellite orbit against the octupole potential $\Phi_3$. 

\noindent(iv) The quadrupole potential associated with $J_2$ is the
strongest non-Keplerian potential for satellites in low Earth orbit by
several orders of magnitude. Let us then write the perturbing
potential is $\la\Phi\ra=\la\Phi_2\ra + \epsilon \la\phi\ra$ where
$\epsilon\ll1$ and $\phi(\bfr)$ represents the potentials from higher
order multipoles, the Moon and Sun, etc. Since $\la\Phi_2\ra_{,22}=0$
(see paragraph [i]) we have to first order in $\epsilon$
\begin{align}
\lambda&=\pm {2i\over\sqrt{GMa}} \nonumber \\
&\quad \times \left[\la\Phi_2\ra_{,1}^2 +\epsilon\la\Phi_2\ra_{,1}\big(2\la\phi\ra_{,1}+\half\la\phi\ra_{,22}\sin^2\beta\big)\right]^{1/2}.
\label{eq:gggr}
\end{align}
At most inclinations $\la\Phi_2\ra_{,1}^2$ is much larger than the
other term since the latter is multiplied by $\epsilon\ll1$; thus
$\lambda$ is pure imaginary and the orbit is stable. Physically, the
rapid precession of the angular-momentum and eccentricity vectors
averages out the perturbations from other sources and
suppresses the instabilities that they would otherwise induce. However,
$\la\Phi_2\ra_{,1}^2$ is proportional to $(1-5\cos^2\beta)$ so near
the critical inclination $\beta_{\rm crit}$ there can be a
narrow range of inclinations in which the square brackets in Eq.\ 
(\ref{eq:gggr}) are negative  so the orbit is unstable. 

\section{Lidov--Kozai oscillations}

\label{sec:lidov}

\noindent
The nonlinear trajectories of the linear instabilities we have
described are known as Kozai, Kozai--Lidov, or Lidov--Kozai
oscillations. Although Laplace had all of the tools needed to
investigate this phenomenon, it was only discovered in the early
1960s by Lidov\cite{lid61}
in the Soviet Union and brought to the West by Kozai\cite{kozai62}. The
simplest, and most astrophysically relevant, examples of Lidov--Kozai
oscillations arise when a distant third body perturbs a binary system,
as in the discussion of the lunar potential in \S\ref{sec:moon}. The
perturbing potential $\la\Phimoon\ra$ (Eq.\ \ref{eq:ejdef}) depends on
the orbital elements of the satellite through the semi-major axis $a$,
the eccentricity $e$, and the projections of the eccentricity and
angular-momentum vectors on the lunar orbit axis, $\bfe\cdot\bfnmoon$
and $\bfj\cdot\bfnmoon$.  The semi-major axis is conserved because we
have orbit-averaged (see discussion in \S\ref{sec:secular multipole}),
$\bfj\cdot\bfnmoon$ is conserved because the orbit-averaged potential
is axisymmetric, and the Hamiltonian $\la\Phimoon\ra$ is conserved
because it is autonomous. Given these four variables and three
conserved quantities, the orbit-averaged oscillation has only one
degree of freedom and thus is integrable,\citep{kozai62,kn07} although the integrability
disappears when higher-order multipole moments are included or the
angular momentum in the inner orbit is not small compared to the outer
orbit.\citep{fkr00,kdm11,naoz13}

Some properties of the oscillations are straightforward to
determine. Since $\la\Phimoon\ra$ and $\bfj\cdot\bfnmoon$ are
conserved, Eq.\ (\ref{eq:ejdef}) tells us that the eccentricity $e$
and the normal component of the eccentricity $\bfe\cdot\bfnmoon$ must
evolve along the track $5(\bfe\cdot\bfnmoon)^2
-2e^2=\mbox{\,constant}$.  If the satellite is initially on a circular
orbit, the constant must be zero so
$(\bfe\cdot\bfnmoon)^2=\frac{2}{5}e^2$. Since $\bfe\cdot\bfj=0$,
$(\bfe\cdot\bfnmoon)^2/e^2$ cannot exceed $\sin^2\beta$ where $\beta$
is the inclination, so $\sin^2\beta\ge\frac{2}{5}$ if $e>0$, which
immediately gives the stability criterion for circular orbits,
$\beta<\beta_{\rm L}$ (Eq.\ \ref{eq:betacrit}). If the inclination
$\beta_0$ of the initial circular orbit exceeds $\beta_{\rm L}$, then
$(\bfj\cdot\bfnmoon)^2=(1-e^2)\cos^2\beta=\cos^2\beta_0$ and we find
$e^2\le 1-\frac{5}{3}\cos^2\beta_0$ which gives the maximum
eccentricity achieved in the Lidov--Kozai oscillation. At the maximum
eccentricity the inclination is $\beta_{\rm L}$.

As an example, Lidov pointed out that if the inclination of the lunar
orbit to the ecliptic were changed to $90^\circ$, with all other
orbital elements kept the same, this ``vertical Moon'' would collide
with the Earth in about four years as a result of a Lidov--Kozai
oscillation induced by the gravitational field of the Sun. More recent
work shows that Lidov--Kozai oscillations may play a significant role
in the formation and evolution of a wide variety of astrophysical
systems. These include:

\begin{itemize}

\item The giant planets in the solar system are surrounded by over 100
  small ``irregular'' satellites, most of them discovered in the last
  two decades. These satellites, typically $<100$ km in radius, are
  found at much larger semi-major axes and have more eccentric and
  inclined orbits than the classical satellites. One of the striking
  features of their orbital distribution is that no satellites are
  found with inclinations between $55^\circ$ and $130^\circ$ (relative
  to the ecliptic), although prograde orbits with smaller inclination
  and retrograde orbits with larger inclination are common ($\sim
  20\%$ and $\sim 80\%$ of the total population, respectively). This
  gap is explained naturally by Lidov--Kozai oscillations:
  at inclinations close to $90^\circ$ the oscillations are so strong
  that the satellite either collides with one of the much larger
  classical satellites or the planet at periapsis, or escapes from
  the planet's gravitational field at apapsis.\citep{car02,nes03}

\item Some extrasolar planets have remarkably high
  eccentricities---the current record-holder has $e=0.97$---and it is
  likely that some or most of the highest eccentricities have been
  excited by Lidov--Kozai oscillations due to a companion star\citep{htt97,inn97,maz97}. In
  fact, the four planets with the largest known eccentricities {\it
    all} orbit host stars with companions.\citep{wri11}

\item ``Hot Jupiters'' are giant planets orbiting within 0.1 AU of
  their host star, several hundred times closer than the giant planets
  in our own solar system. Such planets cannot form {\it in situ}. One
  plausible formation mechanism is ``high-eccentricity migration'',
  which involves the following steps: the planet forms at 5--10 AU
  from the host star, like Jupiter and Saturn in our own solar system;
  it is excited to high eccentricity by gravitational scattering off
  another planet; Lidov--Kozai oscillations due to a companion star or
  other giant planets periodically bring the planet so close to the
  host star that it loses orbital energy through tidal friction; the
  orbit then decays, at a faster and faster rate,
  until the planet settles into a circular orbit close to the host
  star.\citep{wm03,ft07}

\item Many stars are found in close binary systems, with separations
  of only a few stellar radii. Forming such systems is a challenge,
  since the radius of a star shrinks by a large factor during its
  early life. It is possible that most or all close binary systems
  were formed from much wider binaries by the combined effects of
  Lidov--Kozai oscillations induced by a distant companion star and
  tidal friction.\citep{ft07} Supporting this hypothesis is the
  remarkable observation that almost all close binary systems (orbital
  period less than 3 days) have a tertiary companion
  star.\citep{tok06} The formation rate of Type Ia supernovae may be
  dominated by a similar process in triple systems containing a white
  dwarf-white dwarf binary and a distant companion: either
  Lidov--Kozai oscillations plus energy loss through gravitational
  radiation\citep{tho11} or Lidov--Kozai oscillations that excite the
  binary to sufficiently high eccentricity that the two white dwarfs
  collide.\cite{kus13}

\item Most galaxies contain supermassive black holes at the centers,
  and when galaxies merge their black holes spiral towards the center
  of the merged galaxy through dynamical friction.\citep{bbr80}
  However, dynamical friction becomes less effective once the black
  holes have formed a tightly bound binary, and it is unclear whether
  the black hole inspiral will ``stall'' before gravitational
  radiation becomes effective. Lidov--Kozai oscillations in the
  binary, induced either by the overall gravitational field of the
  galaxy or the presence of a third black hole, can pump
  the binary to high eccentricity where gravitational radiation is
  more efficient, leading to inspiral and merger of the black
  holes.\citep{bls02}

\end{itemize}

\section{Discussion}

\noindent
Satellites in low Earth orbit are subject to a variety of
non-Keplerian perturbations, both from the multipole moments of the
Earth's gravitational field and from tidal forces from the Moon and
Sun. For practical purposes in astrodynamics, the effects of these
perturbations have been well-understood for decades. Nevertheless, in
generic perturbing potentials---even axisymmetric ones---many orbits
are likely to be unstable and crash in a short time; a classic example
is Lidov's vertical lunar orbit, described in \S\ref{sec:lidov}. Thus
it is worthwhile to understand physically what features
the perturbing potential must have so that low Earth orbits are
stable.

Our most general result, for an axisymmetric perturbing potential, is
that the orbit is stable if (Eq.\ \ref{eq:ggg})
\begin{equation}
\frac{\p\la\Phi\ra}{\p e^2}\left(\frac{\p\la\Phi\ra}{\p
    e^2}+\frac{1}{2}\frac{\p^2\la\Phi\ra}{\p (\bfe\cdot\bfz)^2}\sin^2\beta\right)
> 0,
\label{eq:ggga}
\end{equation}
where $\bfz$ is the symmetry axis of the potential and $\la\Phi\ra$ is
the orbit-averaged potential written as a function of $e^2$,
$\bfe\cdot\bfz$, and $(\bfj\cdot\bfz)^2=\cos^2\beta$ with $\beta$ the inclination. It is an
accidental property of the potential due to an internal quadrupole
moment that the coefficient of $\sin^2\beta$ vanishes and the motion
is always stable. This statement is not meant to imply
that all other potentials are unstable---the $J_4$ potential analyzed
in \S\ref{sec:jfour} is unstable only over a total range of
inclinations of about $8^\circ$, and Figure \ref{fig:one} shows that
other potentials can be stable for most inclinations---but the
potential associated with $J_2$ is the only axisymmetric potential
arising from a physically plausible mass distribution for which {\it
  all} inclinations are stable. 

We have also shown that because the other perturbing potentials are
much smaller than the one due to the Earth's quadrupole, all low Earth
orbits are stable except perhaps in a small region near the critical
inclination $\beta_{\rm crit}=\cos^{-1}\sqrt{1/5}=63.43^\circ$, where
$\p\la\Phi\ra/\p e^2=0$ for the quadrupole potential.

We chose to investigate only azimuthally symmetric terms in the
perturbing potential, but we strongly suspect that similar arguments
apply to small non-axisymmetric potentials: the strong quadrupole
potential of the Earth guarantees stability except for small
interval(s) in the inclination $\beta$. These intervals will be centered on
resonances\citep{vas74,pal07,kd11} between the precession frequencies of the angular-momentum
vector and the eccentricity vector (eqs.\ \ref{eq:omdef} and
\ref{eq:lambda}), that is, when $1-5\cos^2\beta=2n\cos\beta$ for
integer $n$
($\beta=46.38^\circ,63.43^\circ,73.14^\circ,78.46^\circ,\cdots$). 

Finally, we hope that this paper will introduce readers to the use of
vector elements to study the long-term evolution of Keplerian
orbits. These elements are not new\citep{mil39,mus54,ttn09}
but they deserve to be more widely known.

We thank Boaz Katz and Renu Malhotra for comments and insights. 

\bibliographystyle{phaip}

\bibliography{yavetz}

\begin{thebibliography}{10}

\bibitem{king-hele1958}
D.~G. {King--Hele},
\newblock Proceedings of the Royal Society of London, Series A {\bf 247}, 49
  (1958).

\bibitem{el'yasberg1965}
P.~E. {El'yasberg},
\newblock {\em {Introduction to the Theory of Flight of Artificial Earth
  Satellites}},
\newblock Moscow: Izdatel'stvo Nauka Glavnaya Redaktsiya. Translated by the
  Israel Program for Scientific Translations, 1965.

\bibitem{wgs2000}
N.~K. {Pavlis}, S.~A. {Holmes}, S.~C. {Kenyon}, and J.~K. {Factor},
\newblock Journal of Geophysical Research: Solid Earth {\bf 117}, B04406
  (2012).

\bibitem{luzum2011}
B.~{Luzum} et~al.,
\newblock Celestial Mechanics and Dynamical Astronomy {\bf 110}, 293 (2011).

\bibitem{mil39}
M.~{Milankovich},
\newblock Bull. Serb. Acad. Math. Nat. A {\bf 6}, 1 (1939).

\bibitem{ttn09}
S.~{Tremaine}, J.~{Touma}, and F.~{Namouni},
\newblock Astron.\ J. {\bf 137}, 3706 (2009).

\bibitem{crit}
{See for example Garfinkel, B.\ 1960, Astron,\ J., 65, 624; Lubowe, A.~G.\
  1969, Celestial Mechanics, 1, 6}.

\bibitem{lid61}
M.~L. {Lidov},
\newblock {The approximate analysis of the evolution of the orbits of
  artificial satellites},
\newblock in {\em {Problems of the Motion of Artificial Celestial Bodies}},
  pages 119--134, Academy of Sciences, Moscow, 1961.

\bibitem{kd11}
B.~{Katz} and S.~{Dong},
\newblock {arXiv:1105.3953 (2011)}.

\bibitem{kozai62}
Y.~{Kozai},
\newblock Astron.\ J. {\bf 67}, 591 (1962).

\bibitem{kn07}
H.~{Kinoshita} and H.~{Nakai},
\newblock Celestial Mechanics and Dynamical Astronomy {\bf 98}, 67 (2007).

\bibitem{fkr00}
E.~B. {Ford}, B.~{Kozinsky}, and F.~A. {Rasio},
\newblock \apj {\bf 535}, 385 (2000).

\bibitem{kdm11}
B.~{Katz}, S.~{Dong}, and R.~{Malhotra},
\newblock Physical Review Letters {\bf 107}, 181101 (2011).

\bibitem{naoz13}
S.~{Naoz}, W.~M. {Farr}, Y.~{Lithwick}, F.~A. {Rasio}, and J.~{Teyssandier},
\newblock Mon.\ Not.\ Royal.\ Astron.\ Soc. {\bf 431}, 2155 (2013).

\bibitem{car02}
V.~{Carruba}, J.~A. {Burns}, P.~D. {Nicholson}, and B.~J. {Gladman},
\newblock Icarus {\bf 158}, 434 (2002).

\bibitem{nes03}
D.~{Nesvorn{\'y}}, J.~L.~A. {Alvarellos}, L.~{Dones}, and H.~F. {Levison},
\newblock Astron.\ J. {\bf 126}, 398 (2003).

\bibitem{htt97}
M.~{Holman}, J.~{Touma}, and S.~{Tremaine},
\newblock Nature {\bf 386}, 254 (1997).

\bibitem{inn97}
K.~A. {Innanen}, J.~Q. {Zheng}, S.~{Mikkola}, and M.~J. {Valtonen},
\newblock Astron.\ J. {\bf 113}, 1915 (1997).

\bibitem{maz97}
T.~{Mazeh}, Y.~{Krymolowski}, and G.~{Rosenfeld},
\newblock Astrophys.\ J. {\bf 477}, L103 (1997).

\bibitem{wri11}
J.~T. {Wright} et~al.,
\newblock Publ.\ Astr.\ Soc.\ Pacific {\bf 123}, 412 (2011).

\bibitem{wm03}
Y.~{Wu} and N.~{Murray},
\newblock Astrophys.\ J. {\bf 589}, 605 (2003).

\bibitem{ft07}
D.~{Fabrycky} and S.~{Tremaine},
\newblock Astrophys.\ J. {\bf 669}, 1298 (2007).

\bibitem{tok06}
A.~{Tokovinin}, S.~{Thomas}, M.~{Sterzik}, and S.~{Udry},
\newblock A \& A {\bf 450}, 681 (2006).

\bibitem{tho11}
T.~A. {Thompson},
\newblock Astrophys.\ J. {\bf 741}, 82 (2011).

\bibitem{kus13}
D.~{Kushnir}, B.~{Katz}, S.~{Dong}, E.~{Livne}, and R.~{Fern{\'a}ndez},
\newblock (2013).

\bibitem{bbr80}
M.~C. {Begelman}, R.~D. {Blandford}, and M.~J. {Rees},
\newblock Nature {\bf 287}, 307 (1980).

\bibitem{bls02}
O.~{Blaes}, M.~H. {Lee}, and A.~{Socrates},
\newblock Astrophys.\ J. {\bf 578}, 775 (2002).

\bibitem{vas74}
M.~A. {Vashkov'yak},
\newblock Cosmic Research {\bf 12}, 757 (1974).

\bibitem{pal07}
J.~F. {Palaci{\'a}n},
\newblock Celestial Mechanics and Dynamical Astronomy {\bf 98}, 219 (2007).

\bibitem{mus54}
P.~{Musen},
\newblock Astron.\ J. {\bf 59}, 262 (1954).

\end{thebibliography}

\end{document}